\documentclass[english,american,preprint,aps,prl,ams,multicols]{revtex4}
\usepackage[T1]{fontenc}
\usepackage[latin1]{inputenc}
\usepackage{amssymb}

\makeatletter

\providecommand{\tabularnewline}{\\}

\usepackage{ae}
\usepackage{aecompl}
\usepackage{amsmath}
\usepackage{graphicx}
\usepackage{epstopdf}

\usepackage{babel}
\makeatother
\begin{document}

\title{Quantum teleportation between nanomechanical modes}

\author{L. Tian \& S. M. Carr}

\affiliation{Atomic Physics Division, National Institute of Standards and Technology, 
100 Bureau Drive, Stop 8423, Gaithersburg, Maryland 20899, USA}

\date{\today}

\begin{abstract}
We study a quantum teleportation scheme between two nanomechanical
modes without local interaction. The nanomechanical modes are linearly
coupled to and connected by the continuous variable modes of a superconducting
circuit consisting of a transmission line and Josephson junctions. We calculate
the fidelity of transferring Gaussian states at finite temperature
and non-unit detector efficiency.  For coherent state, a fidelity above the
classical limit of $1/2$ can be achieved for a large range of parameters.
\end{abstract}
\maketitle

Stimulating progress has been made in observing the mechanical vibration of nanostructures
towards the quantum limit.  The flexural mode of micro-fabricated resonator was
recently measured by a superconducting single electron transistor (SSET) with a resolution
approaching the Heisenberg limit \cite{schwabScience2004}.  Experimental realization 
\cite{RoukesNature} of nanomechanical oscillators with resonant frequencies 
$\gtrsim$ 1 GHz  enables novel coupling with superconducting quantum devices.  It has been proposed that a piezoelectric nanomechanical
resonator, with experimentally demonstrated high frequency and high quality factor, be used
to perform controlled quantum logic gates between Josephson junction
qubits \cite{ClelandGellerPRL}.  Considering these results, the prospects are promising for 
the observation and application of quantum behavior in nanoelectromechanical systems 
\cite{BlencowePhysRep2004,schwabPRL2002}.

The nanomechanical modes, as a macroscopic continuous variable system
with high quality factor, provide a unique system for the storage,
local manipulation, and transfer of quantum information.
Among such applications, one fundamental element is quantum teleportation
\cite{bennettPRL93,VaidmanPRA,braunsteinPRL98a},
which has been demonstrated with various microscopic systems such
as photons and trapped ions \cite{bouwmeesterNature97,iontele}. Previously, it was 
proposed that the entanglement and teleportation
be achieved between an optical and a nanomechanical mode of a
mirror via pondermotive forces \cite{bouwmeesterPRL03,tombesiPRL02,tombesiPRL03}.

In this paper, we study a practical scheme of quantum teleportation
between nanomechanical modes in an all solid-state circuit consisting of a
superconducting transmission line and Josephson junctions. Each nanomechanical
mode couples locally with the superconducting phase variable of the junctions, which
can be treated as a continuous variable mode when choosing the Josephson
energy to be much larger than the charging energy, in contrast to
the qubit model in Ref.\,\cite{ClelandGellerPRL}. The transmission
line acts as a dynamical connection between the phase variables at
distant locations and hence forms an effective media, the ``vacuum''
in its optical counter part, between the nanomechanical modes. The superconducting 
transmission line has recently been shown
to couple strongly with superconducting qubits \cite{yaleCQEDexp}
and has been proposed as a tool to interface different systems
for scalable quantum computing \cite{TianInterfacingPRL2004}. 
After deriving the coupling constants for our scheme, we calculate
the fidelity by a Wigner function approach when considering the effects of temperature 
and detector noise.  We use this approach to derive the parameter regime in which
a fidelity of transferring coherent state above the classical limit \cite{braunsteinquantclassPRA01} 
of $F_{c}=1/2$ can be achieved. 

\begin{figure}
 \includegraphics[bb=60bp 150bp 576bp 632bp,clip,width=7cm]{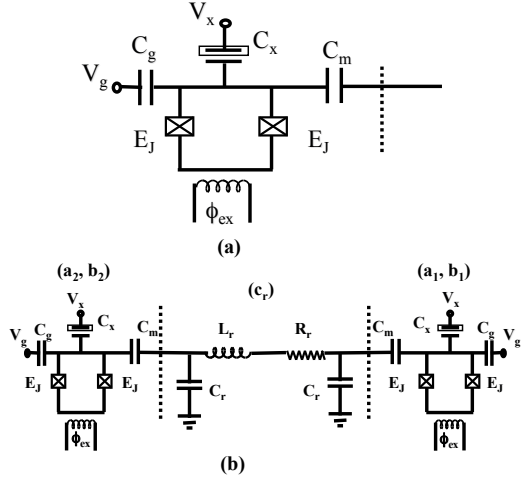}
\caption{Circuit.
(a) Coupled nanomechanical resonator and superconducting island. The resonator
is drawn as the boxed capacitor plate. (b) The circuit for quantum teleportation
includes locally coupled modes $(a_{1},b_{1})$ and $(a_{2},b_{2})$ and the
transmission line mode $c_{r}$.}
\label{figure1}
\end{figure}

The central element in our scheme consists of two capacitavely-coupled continuous variable modes:
the vibration of a nanomechanical resonator and the gauge invariant phase of a superconducting island
connected to two Josephson junctions, shown in Fig.\ref{figure1}(a).
The junctions have Josephson energy $E_{J}$ and capacitance $C_{J}$
and can be controlled by the gate voltage $V_{g}$ via a gate capacitance
$C_{g}$ and the external flux $\phi_{ex}$ in the superconducting
loop. The nanomechanical resonator is biased at the voltage $V_{x}$
and couples with the superconducting island via a capacitance $C_{x}\approx C_{x}^{0}(1+\hat{x}/d_{0})$ 
that depends on the canonical displacement $\hat{x}$ of the mechanical
vibration and a characteristic distance $d_{0}$ between the island
and the resonator.

The Hamiltonian of the system in Fig.\ref{figure1}(a) is $H_{t}=H_{\varphi}+H_{nr}+H_{int}$.
The first term \begin{equation}
H_{\varphi}=\frac{\left(\hat{p}_{\varphi}+Q_{0}\right)^{2}}{2C_{\Sigma}}-E_{J}^{eff}\cos\hat{\varphi},\label{eq:Hphi}\end{equation}
 describes the superconducting mode with the total capacitance $C_{\Sigma}=2C_{J}+C_{g}+C_{x}^{0}+C_{m}$
(see Fig.\ref{figure1}(b) for $C_{m}$), $Q_{0}=C_{x}^{0}V_{x}+C_{g}V_{g}$
and the effective Josephson energy $E_{J}^{eff}=2E_{J}\cos\phi_{ex}$.
In the regime \cite{LupascuPRL2004,LeeIEEE2005} $E_{J}^{eff}\gg E_{c}$
with $E_{C}=e^{2}/2C_{\Sigma}$, the superconducting mode can be described
as an harmonic oscillator with the frequency $\omega_{\varphi}=\sqrt{8E_{J}^{eff}E_{C}}/\hbar$
, i.e. $H_{\varphi}=\hbar\omega_{\varphi}\hat{b}^{\dagger}\hat{b}$.
Below we assume $Q_{0}=0$. For the resonator mode, $H_{nr}=\hbar\omega_{nr}\hat{a}^{\dagger}\hat{a}$
with frequency $\omega_{nr}$ amd mass $m_{nr}$. Here $\hat{a}$
($\hat{b}$) and $\hat{a}^{\dagger}$($\hat{b}^{\dagger}$) are the
annihilation and creation operators of the resonator mode (superconducting
mode) respectively. For the bias voltage $2V_{x}^{0}\cos\omega_{d}t$
with amplitude $V_{x}^{0}$ and driving frequency $\omega_{d}$, the
coupling between these two modes can be derived as $H_{int}=i\lambda_{ab}\cos\omega_{d}t(\hat{a}-\hat{a}^{\dagger})(\hat{b}+\hat{b}^{\dagger})$
with the coupling constant  
\begin{equation}
\lambda_{ab}=\frac{2C_{x}^{0}V_{x}^{0}}{e}\frac{\delta x_{0}}{d_{0}}\left(2E_{J}^{eff}\right)^{1/4}E_{C}^{3/4}\label{eq:lamb_ab}
\end{equation}
where $\delta x_{0}=\sqrt{\hbar/2m_{nr}\omega_{nr}}$. 

The basic operations of a continuous variable quantum teleportation
scheme can be performed by the coupling $H_{int}$. In the interaction
picture, with the driving frequency $\omega_{d}=\omega_{\varphi}+\omega_{nr}$, 
the coupling is $H_{I}=i\lambda_{ab}\left(\hat{a}\hat{b}-\hat{a}^{\dagger}\hat{b}^{\dagger}\right)$
in the rotating wave approximation, which generates squeezing and
entanglement between the modes. With $\omega_{d}=\omega_{\varphi}-\omega_{nr}$,
$H_{I}=i\lambda_{ab}\left(\hat{a}\hat{b}^{\dagger}-\hat{a}^{\dagger}\hat{b}\right)$,
which generates a $50$-$50$ beam splitter operation when applied
for a duration of $\pi/4\lambda_{ab}$ and generates a swapping operation
when applied for a duration of $\pi/2\lambda_{ab}$.

An unknown state in the nanomechanical mode $a_{1}$ can be teleported
to another nanomechanical mode $a_{2}$ at a remote location via the
superconducting network shown in Fig.\ref{figure1}(b). The crucial
steps in this scheme are to first generate entanglement between the 
locally coupled modes $a_{2}$ and $b_{2}$ 
and then transfer the entanglement via the superconducting transmission
line to the superconducting mode $b_{1}$ that couples locally with $a_{1}$. 
The transmission line mode can be treated as a LC oscillator
with frequency $\omega_{r}$ ($<\omega_{\varphi}$) and couples with
the superconducting modes as $H_{IR}=\lambda_{bc}\left(\hat{b}_{i}\hat{c}_{r}^{\dagger}-\hat{c}_{r}^{\dagger}\hat{b}_{i}\right)$
with the coupling constant
\begin{equation}
\lambda_{bc}=\hbar\sqrt{\frac{C_{m}^{2}\omega_{\varphi}\omega_{r}}{4C_{\Sigma}C_{r}}}\label{eq:lamb_bc}
\end{equation}
with $C_{r}\gg C_{m}$. Here $\hat{c}_{r}$ and $\hat{c}_{r}^{\dagger}$
are creation and annihilation operators for the transmission line mode
and $C_{m}$ is the coupling capacitance. 

In Table \ref{table1}, the steps of the teleportation process are shown starting from 
the generation of entanglement between $a_{2}$ and $b_{2}$.  Then set $V_{x}^{0}=0$  to turn off the coupling between $a_{2}$ and $b_{2}$ and tune the frequency 
of $b_{2}$ to be in resonance with $\omega_{r}$
by adjusting the flux $\phi_{ex}$ to achieve the swapping between
$b_{2}$ and $c_{r}$; and similarly between the modes $b_{1}$ and
$c_{r}$. An entanglement between the distant modes $a_{2}$ and
$b_{1}$ is now formed. The standard steps of a quantum teleportation
scheme \cite{braunsteinPRL98a} can then be applied. The displacement \cite{TianEntangle2005}
of $a_{2}$ at the end of the scheme can be achieved by displacing the mode $b_{2}$
with a time dependent gate charge $Q_{0}$ followed by mixing $a_{2}$
and $b_{2}$ with a beam splitter operation.

The scheme discussed above requires 
\begin{equation}
\frac{\omega_{nr}}{Q},\gamma_{d}\ll\frac{\lambda_{ab}}{\hbar},\frac{\lambda_{bc}}{\hbar},\tau_{m}^{-1}\ll\omega_{nr},\omega_{r},\omega_{\varphi}\ll E_{J}^{eff}\label{eqcc}
\end{equation}
where $Q$ is the quality factor of the resonator and $\gamma_{d}$
is the dissipation rate of the superconducting modes. The first inequality
ensures that the scheme is not affected by the dissipation from the
environment. For the rotating wave approximation to be valid,
the second inequality needs to be satisfied. The last inequality ensures
that the harmonic oscillator model of the phase variables is valid.
We choose the parameters of the junctions: $C_{\Sigma}\approx0.1\:\textrm{pF}$
and $E_{J}=500\,\textrm{GHz}$ \cite{LupascuPRL2004,LeeIEEE2005},
which gives $\omega_{\varphi}\approx28\,\textrm{GHz}$. For the resonators \cite{schwabScience2004,RoukesNature},
$\omega_{nr}\approx1\,\textrm{GHz}$ and $\delta x_{0}=22\,\textrm{fm}$.
With $d_{0}=100\,\textrm{nm}$, $C_{x}^{0}=0.65\,\textrm{fF}$ and
$V_{x}^{0}=2.4\,\textrm{V},$ we derive the coupling constant $\lambda_{ab}=5\,\textrm{MHz}$.
For the transmissition line \cite{yaleCQEDexp}, $C_{m}=1\,\textrm{fF}$,
$C_{r}=4\,\textrm{fF}$ and $\omega_{r}\approx5\,\textrm{GHz}$, resulting in the
coupling constant $\lambda_{bc}=0.3\,\textrm{GHz}$. Experimentally, quality factors of 
$Q \approx10^{3}$ have been demonstrated and thus Eq. (\ref{eqcc}) is satisfied.
Assuming a larger quality factor, for example $Q=10^{4}$, it has a negligible effect on the scheme. For the superconducting modes with our parameters, the dissipation has been
measured to be $\sim100\,\textrm{kHz}$ \cite{LupascuPRL2004,LeeIEEE2005}
and easily satisfies Eq. (\ref{eqcc}).  The measurement time $\tau_{m}=50\,\textrm{ns}$
listed in Table \ref{table1} satisfies the condition in Eq. (\ref{eqcc}).
\begin{table}
\begin{tabular}{|c|c|c|}
\hline 
operation&
 modes&
 duration \tabularnewline
\hline
squeezing with $r_{2}=1.5$&
 ($a_{2},b_{2}$)&
 $48\,\textrm{ns}$\tabularnewline
\hline
swapping&
 ($b_{2},c_{r}$)&
 $0.8\:\textrm{ns}$\tabularnewline
\hline
swapping&
 ($c_{r},b_{1}$)&
 $0.8\,\textrm{ns}$\tabularnewline
\hline
$50$-$50$ beam splitter&
 ($a_{1},b_{1}$)&
 $25\,\textrm{ns}$\tabularnewline
\hline
Bell-measurement $\delta_{x},\delta_{p}$&
 ($a_{2},b_{2}$)&
 $50\,\textrm{ns}$ \tabularnewline
\hline
\end{tabular}

\caption{\label{table1} List of the operations in the quantum teleportation
scheme. The modes involved in each operation are shown in the second
column. The durations of the operations are estimated with the parameters
given in the text.}
\end{table}

In a solid-state system and the microwave regime, noise and finite
temperature can seriously affect the fidelity of the teleportation.
We study these effects with a Wigner function approach. The modes
$i=b_{1},\, b_{2},\, c_{r},\, a_{2}$ are initially in thermal equilibrium
labeled by a dimensionless constant $\Theta_{i}=\coth\frac{\hbar\omega_{i}}{2k_{B}T}$
at temperature $T$. In our scheme, $\omega_{\varphi},\omega_{r}\gg k_{B}T$
so that $\Theta_{bi,r}=1$; while $\omega_{nr}\sim k_{B}T$, we have
$\Theta_{ai}\geq1$. The squeezing operation applied for a time $t_{s}$
(and hence a squeezing parameter $r_{2}=\lambda_{ab}t_{s}$) generates
entanglement \cite{SimonSeparabilityPRL2000,DuanSeparabilityPRL2000}
with the Wigner function $W^{sq}(x_{a2},p_{a2},x_{b2},p_{b2})$. A
measurement on variable $x_{i}$ (or similarly on $p_{i}$) can be
described as a POVM $\Pi_{x_{m}}(x_{i})\propto e^{-\frac{(x_{i}-x_{m})^{2}}{2\delta_{x}^{2}}}$
with the noise amplitude $\delta_{x}$ (or $\delta_{p}$) and measurement
record $x_{m}$ ($p_{m}$). 

For an initial state $W^{in}(x_{a1},p_{a1})$, after displacing the
mode $a_{2}$ by $(\sqrt{2}x_{m},\sqrt{2}p_{m})$ and averaging over
the measurement record, the final Wigner function is
\begin{widetext}
\begin{eqnarray}
W^{f}(x_{a2},p_{a2}) & = & C\int d^{2}x_{b1}d^{2}x_{a1}d^{2}x_{m}e^{-\frac{(x_{a1}-x_{m})^{2}}{2\delta_{x}^{2}}-\frac{(p_{b1}-p_{m})^{2}}{2\delta_{p}^{2}}}\nonumber \\
 & \times & W^{in}(x_{b1},p_{a1})W^{sq}(x_{a2}-\sqrt{2}x_{m},p_{a2}-\sqrt{2}p_{m},x_{b1}-\sqrt{2}x_{a1},\sqrt{2}p_{b1}-p_{a1})
\label{eq:wf}
\end{eqnarray}
\end{widetext}
where $C$ is the renormalization factor and $d^{2}x_{i}=dx_{i}dp_{i}$.
It can be shown that $W^{f}=W^{in}\circ G_{\mu}(x_{a2},p_{a2})$ is
a convolution between the initial state and a Gaussian state $G_{\mu}(x,p)=\frac{1}{\pi\sqrt{\mu_{x}\mu_{p}}}e^{-x^{2}/\mu_{x}-p^{2}/\mu_{p}}$
with $\mu=(\mu_{x},\mu_{p})$ and
\begin{equation}
\mu_{i}=4\delta_{i}^{2}+\frac{1}{2}(1+\Theta_{a2})e^{-2r_{2}}\label{eq:mu}
\end{equation}
for $i=x,p$. For a Gaussian state $W^{in}=G_{\mu_{0}}(x_{a1}-x_{in},p_{a1}-p_{in})$,
the final state is the Gaussian state $G_{\mu+\mu_{0}}(x_{a2}-x_{in},p_{a2}-p_{in})$.
At large squeezing, $\mu_{x,p}$ are independent of the dimensionless
temperature $\Theta_{a2}$; the entanglement can suppress the
effect of the temperature, but can not suppress the effect of the detector
noise.

The fidelity defined by $F_{m}(\rho_{a1}^{in})=\left(\textrm{Tr}(\sqrt{\sqrt{\rho_{a1}^{in}}\rho_{a2}^{f}\sqrt{\rho_{a1}^{in}}})\right)^{2}$
describes the overlap between the final state with the density matrix
$\rho_{a2}^{f}$ and the initial state with the density matrix $\rho_{a1}^{in}$.
For a Gaussian initial state with $\mu_{0}=(\frac{\Theta_{a1}}{2},\frac{\Theta_{a1}}{2})$,
we derive \cite{BanFidPRA2004,MarionFidPRL02}: 
\begin{widetext}
\begin{eqnarray}
F_{m}(\rho_{a1}^{in}) & = & \frac{2}{\sqrt{(1+2\Theta_{a1}\sigma_{x})(1+2\Theta_{a1}\sigma_{p})}-\sqrt{(4\sigma_{x}\sigma_{p}-1)(\Theta_{a1}^{2}-1)}},\label{eq:Fmf}
\end{eqnarray}
\end{widetext} 
where $\sigma_i=\mu_i+\Theta_{a1}/2$ for $i=x,p$ and the fidelity is independent of the initial displacement.  With $\mu_{x,p}=0$,
$F_{m}=1$ and agrees with the above discussion. 

\begin{figure}
\includegraphics[bb=72bp 240bp 540bp 606bp,clip,width=6.5cm]{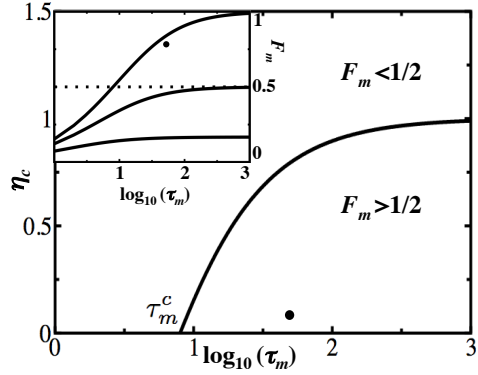}
\caption{Fidelity for transferring
a coherent state.  Main plot: $\eta_{c}(\tau_{m})$ versus $\log_{10}(\tau_{m})$.
The critical measurement time is labeled by $\tau_{m}^{c}$. The solid
dots indicate the operating point with the parameters in Table \ref{table1}.
Inset: $F_{m}$ versus $\tau_{m}$ for $\eta=0,1,5$ respectively
from top to bottom. The dotted line is the classical limit.}
\label{figure2}
\end{figure}

The fidelity of transferring an arbitrary coherent state via a classical
channel is bounded \cite{braunsteinquantclassPRA01} by the upper limit $F_{c}=1/2$.
With quantum entanglement, a fidelity $F_{c} > 1/2$ can be achieved even in
noisy environments.  Following Eq.(\ref{eq:mu}), we study the fidelity
with two parameters: $\eta=\frac{1}{2}(1+\Theta_{a2})e^{-2r_{2}}$
describing the interplay between temperature and squeezing (entanglement),
and the measurement time $\tau_{m}$ which determines the noise amplitude
with $\delta_{x,p}=c_{x,p}/\sqrt{\tau_{m}}$ and $c_{x,p}$ as device-dependent
parameters (below). We find that with a coherent state as the initial
state and at a given $\tau_{m}$, a critical index $\eta_{c}(\tau_{m})$
exists that $F_{m}>1/2$ for $\eta<\eta_{c}$ ; and $F_{m}<1/2$ for
$\eta>\eta_{c}$, as is plotted in Fig.\ref{figure2}. It is obvious
that the effect of finite temperature can be cancelled by the squeezing.
However, for $\tau_{m} < \tau_{m}^{c}$, with $\tau_{m}^{c}$ the critical measurement time
labeled in the figure, 
the fidelity is always less than the classical limit.
With the parameters in Table \ref{table1}, $\eta=0.08$, $\tau_{m}=50\,\textrm{ns}$ 
(as indicated in Fig. \ref{figure2}),
and a temperature of $50\,\textrm{mK}$,
we have the fidelity $F_{m}=0.78$. 

\begin{figure}
\includegraphics[bb=66bp 324bp 534bp 676bp,clip,width=6.5cm]{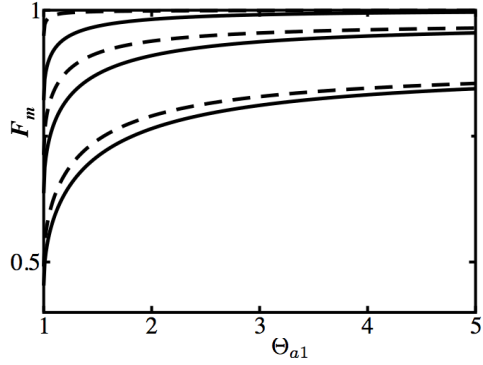}
\caption{Fidelity for transferring
finite temperature state. The solid lines: $\tau_{m}=50\,\textrm{ns}$
and $r_{2}=2,0.5,0$ respectively from top to bottom. The dashed lines:
$\tau_{m}=10^{3}\,\textrm{ns}$ and $r_{2}=2,0.5,0$ respectively
from top to bottom.}
\label{figure3}
\end{figure}

For an initial state at finite temperature with $\Theta_{a1}>1$, the fidelity is plotted in Fig.\ref{figure3}
under the condition $\Theta_{a1}=\Theta_{a2}$. At fixed $r_{2}$ and
$\delta_{x,p} (\tau_{m})$, the fidelity increases
well above $F_{c}=1/2$ with an increase in temperature.
This is because the width of the Wigner function for the initial Gaussian state
increases with temperature and less information is transferred via the teleportation.
%

The Bell-type measurement includes phase sensitive detection of the variables $\hat{x}_{a1}$ and
$\hat{p}_{b1}$.  Here we consider a specific detection scheme using a SET.  
It has been demonstrated that a SET can be operated as a radio-frequency
mixer \cite{clelandMixerAPL2002} that performs
homodyne and heterodyne detections on charge signals and can perform
fast and sensitive measurement \cite{schoelkopfScience1998} of the vibrations of nanomechanical
modes \cite{schwabScience2004,Knobel2003,BlencoweWybourne2000}. In our scheme, the
resonator is biased at a static voltage $V_{x}^{d}$ during the measurement of $\hat{a1}$
and couples with the SET via the capacitance $C_{x}^{d}(\hat{x}_{a1})=C_{x}^{d}(1+\hat{x}_{a1}/d_{1})$
where $d_{1}$ is a normalized distance. The total ac charge signal
is a mixing between the charge of the resonator and an ac gate charge:
$Q=C_{x}^{d}V_{x}^{d}\langle\hat{x}_{a1}\rangle/d_{1}+Q_{ac}\cos\omega_{nr}t$.
Here $\langle\hat{x}_{a1}\rangle=x_{i}\cos\omega_{nr}t+\frac{p_{i}}{m\omega_{a1}}\sin\omega_{nr}t$
with $x_{i}=\langle\hat{x}_{a1}\rangle$ and $p_{i}=\langle\hat{p}_{a1}\rangle$.
Let $\alpha_{g}=\frac{2\pi}{e}Q_{ac}$ and $\alpha_{x}=\frac{2\pi}{ed_{1}}C_{x}^{d}V_{x}^{d}$.
By choosing the static bias to be at the extreme point \cite{clelandMixerAPL2002}, the current response
of the SET is $I_{Q}(t)=I_{0}\cos\frac{2\pi}{e}Q$ with $I_{0}$ the
amplitude of the current. Using the relation: $\cos(\alpha\sin\varphi)=\sum_{n=-\infty}^{\infty}J_{n}(\alpha)\cos n\varphi$
for the Bessel functions $J_{n}$ and $\alpha_{x}x_{i},\,\alpha_{x}\frac{p_{i}}{m\omega_{nr}}\ll\alpha_{g}$,
we derive the static component of the current as
\begin{equation}
I_{0}J_{0}(\alpha_{g})-I_{0}J_{1}(\alpha_{g})\alpha_{x}x_{i},\label{eq:di}
\end{equation}
which is linear in the quadrature $x_{i}$. When choosing $\alpha_{g}=1.9$, we have $J_{1}(1.9)\approx0.58$ and the current
response is maximized. As $J_{n}(x)=-J_{-n}(x)$, the lowest ac component
of the current has the frequency of $2\omega_{nr}$. When the bandwidth of the detector is
below $2\omega_{nr}$, this signal and other ac components are not present in the detection. 
A similar measurement can be conducted on the quadrature variable of $p_{b1}$. 

The sensitivity of the detector is limited by the shot noise of the SET \cite{schoelkopfScience1998}. 
In our scheme, the displacement sensitivity $\sqrt{S_{x}}=\sqrt{S_{q}}/(\partial q/\partial x_{i})$
can be derived as
\begin{equation}
\sqrt{S_{x}}\approx\beta_{hd}\frac{\sqrt{S_{q}}d_{1}}{C_{x}^{d}V_{x}^{d}}.
\label{eq:Sx}
\end{equation}
where $\beta_{hd}\sim1$ is a numerical factor. With a charge sensitivity
of $\sqrt{S_{q}}\sim10^{-6}e/\sqrt{\textrm{Hz}}$ \cite{schoelkopfScience1998},
$C_{x}^{d}=0.5\times10^{-15}\,\textrm{F}$ and $V_{x}^{d}=10\,\textrm{V}$,
this gives $\sqrt{S_{x}}\approx3\times10^{-18}\:\textrm{m}/\sqrt{\textrm{Hz}}$.
At $\tau_{m}=50\,\textrm{ns}$ and our parameters for the resonator,
the accuracy of the displacement is $\Delta x=\sqrt{S_{x}/\tau_{m}}\approx14\,\textrm{fm}$
and $\delta_{x}\approx0.32$. Similarly for the measurement of $\hat{p}_{b1}$,
we can derive the sensitivity of the measurement as $\sqrt{S_{p}}=\beta_{hd}\sqrt{S_{q}}C_{\Sigma}/C_{p}^{d}$.
With $C_{p}^{d}=1\,\textrm{fF}$ and $\tau_{m}=50\,\textrm{ns}$,
we find $\Delta p=0.45\: e$ and $\delta_{p}\approx0.07$.

In conclusion, we studied a quantum teleportation scheme that
transfers information between nanomechanical modes in a solid-state
setup. We showed that at practical parameters, fidelity
above the classical limit can be achieved by utilizing quantum entanglement.

\end{document}